\documentclass{article}
\usepackage{spconf,amsmath,graphicx}
\usepackage{booktabs}
\usepackage{tikz}
\usepackage[font=normalsize]{subfig}
\usepackage{soul}
\usepackage{ulem}
\usepackage{marginnote}

\title{Investigating Disentanglement in a Phoneme-level Speech Codec for Prosody Modeling}
\name{
\begin{tabular}{c}
	Sotirios Karapiperis$^{\star}$,
	Nikolaos Ellinas$^{\star}$,
	Alexandra Vioni$^{\star}$, \\
	Junkwang Oh$^{\dagger}$,
	Gunu Jho$^{\dagger}$,
	Inchul Hwang$^{\dagger}$,
	Spyros Raptis$^{\star}$
\end{tabular}
}

\address{$^{\star}$ Innoetics, Samsung Electronics, Greece \\
	$^{\dagger}$ Mobile eXperience Business, Samsung Electronics, Republic of Korea}

\begin{document}
\ninept
\maketitle
\begin{abstract}
Most of the prevalent approaches in speech prosody modeling rely on learning global style representations or a continuous latent space which encode and transfer the attributes of reference speech.
However, recent work on neural codecs which are based on Residual Vector Quantization (RVQ) already shows great potential offering distinct advantages.
We investigate the prosody modeling capabilities of the discrete space of such an RVQ-VAE model, modifying it to operate on the phoneme-level. We condition both the encoder and decoder of the model on linguistic representations and apply a global speaker embedding in order to factor out both phonetic and speaker information.
We conduct an extensive set of investigations based on subjective experiments and objective measures to show that the phoneme-level discrete latent representations obtained this way achieve a high degree of disentanglement, capturing fine-grained prosodic information that is robust and transferable. The latent space turns out to have interpretable structure with its principal components corresponding to pitch and energy.
\end{abstract}
\begin{keywords}
Prosody Modeling, Speech Synthesis, Vector Quantization, RVQ-VAE
\end{keywords}
\section{Introduction}
\label{sec:intro}
Variational auto-encoders (VAEs) \cite{kingma2013auto} are a popular framework architecture in image and audio generation due to their ability to learn factors of variation of the given dataset in an unsupervised manner. In this paradigm, which has been successfully used in text-to-speech literature in recent years, a VAE operates on speech data shaping a structured latent space. This is coupled with a prior model trained concurrently or separately which learns a mapping between the textual data and the features of this space \cite{kim2021conditional,liu2022controllable,tan2024naturalspeech}.

Vector-quantized VAEs (VQ-VAEs) \cite{van2017neural} offer an extension to this framework by quantizing the latent space, thus forcing the encoder to use a set of learnable code embeddings.
While this bottleneck initially appears to affect reconstruction quality \cite{sun2020generating}, advancements such as Residual Vector Quantization (RVQ) and adversarial training have enabled systems to present state-of-the-art performance in neural audio coding \cite{zeghidour2021soundstream,defossez2022high}.
The choice of a discrete space also facilitates the use of prior Large Language Models (LLMs), which had delivered impressive results in the field of NLP.
At the time of this writing, state-of-the-art approaches mainly use this paradigm, namely VALL-E \cite{wang2023neural} and NaturalSpeech 2 \cite{shen2023naturalspeech}. 

\subsection{Related Work}
\label{sec:related}

In \cite{kim2021conditional,tan2024naturalspeech} a VAE is used with a text-conditioned prior to create a continuous latent space containing both linguistic and acoustic information.
In \cite{liu2022controllable} both the encoder and decoder are conditioned on linguistic information aiming to create a prosodic latent space while concurrently training a prosody-predictor network that learns the mapping between linguistics and this space.
In \cite{van2017neural} the VQ-VAE framework is introduced with the intuition that some modalities, including speech, are better modeled using discrete units. Based on this model, the majority of recent approaches use a discrete space VAE.

Residual Vector Quantization, introduced in SoundStream \cite{zeghidour2021soundstream} and Encodec \cite{defossez2022high}, achieved high compression rates and state-of-the-art reconstruction quality, establishing the term \textit{neural audio codec} in place of VQ-VAE.
The work in \cite{shen2023naturalspeech} extends \cite{tan2024naturalspeech} to operate on the latent space of a neural codec using a diffusion model.
Ideas from NLP, specifically LLMs, have been adapted to the speech synthesis domain by \cite{wang2023neural,borsos2023audiolm,borsos2023soundstorm}. There, the neural codec is used as a speech tokenizer and an autoregressive or parallel transformer-based model is trained to predict the speech tokens from the text.

The modeling of prosody has been approached using various methods. In \cite{wang2018style} global learnable style embeddings are used to capture the variance in speech, whereas \cite{ren2020fastspeech} uses a \emph{variance adaptor} module to predict \emph{local} frame-level prosodic attributes, and \cite{min2021meta,li2022styletts} propose a combination of both approaches.
In contrast with these, VAE-based approaches have been proposed to capture in an unsupervised manner the latent global~\cite{hsu2018hierarchical} or phone-level~\cite{sun2020fully,lee2019robust,liu2022controllable} prosodic attributes. Later works extend these approaches through VQ-VAEs \cite{sun2020generating} and neural codecs \cite{jiang2023mega}.

\subsection{Proposed Method}
\label{sec:proposed}

In prosody modeling, mel encoders have been extensively used to provide trainable global representations of the prosodic\slash acoustic attributes of reference speech~\cite{wang2018style,min2021meta,li2022styletts}.
More fine-grained approaches have also been proposed which employ a text-conditioned VAE to produce phoneme-level prosody representations~\cite{sun2020fully,liu2022controllable}.
However, while increasing research has been directed into employing such models, only relatively limited work has been devoted to analyzing the properties of their latent space and the representations formed therein, and that work is mainly focused on models whose latent representation encodes both linguistic and acoustic information.
Some examples are \cite{williams2021exploring,fong2021analysing} who performed a set of qualitative experiments to investigate the extent to which the different combinations of codes produce intelligible and natural speech.

We propose a phoneme-level speech neural codec specifically trained to model prosodic information in a way that is disentangled from linguistic content and from speaker characteristics. This is achieved by modeling the linguistic information and the speaker characteristics separately and feeding them to the encoder/decoder of the model, thus explaining them away from the latent representation. We train this codec model on multi-speaker data and perform a range of experiments to verify that the representations formed in its latent space are indeed independent of both linguistic and speaker information and that they are tailored to capturing the prosodic aspects of the reference speech.
We further explore this latent space by attempting to explain its structure and properties, and by investigating its ability for prosody modeling and control using both subjective and objective metrics on selected tasks that relate to intelligibility, cross-resynthesis and prosody transfer.

Compared with other speech codec models in the literature, the proposed model has significantly reduced dimensions and complexity while maintaining a very high quality of generated speech. Moreover, its latent space is much more compact and more amenable to interpretation and control. Besides making the model more efficient both during training and at inference, this also suggests that the task of training priors to predict latent codes directly from phonemes may be significantly simplified, a postulation that is left to future research.

\begin{figure}[htb]
	
	\centering
	\centerline{\includegraphics[width=1\linewidth]{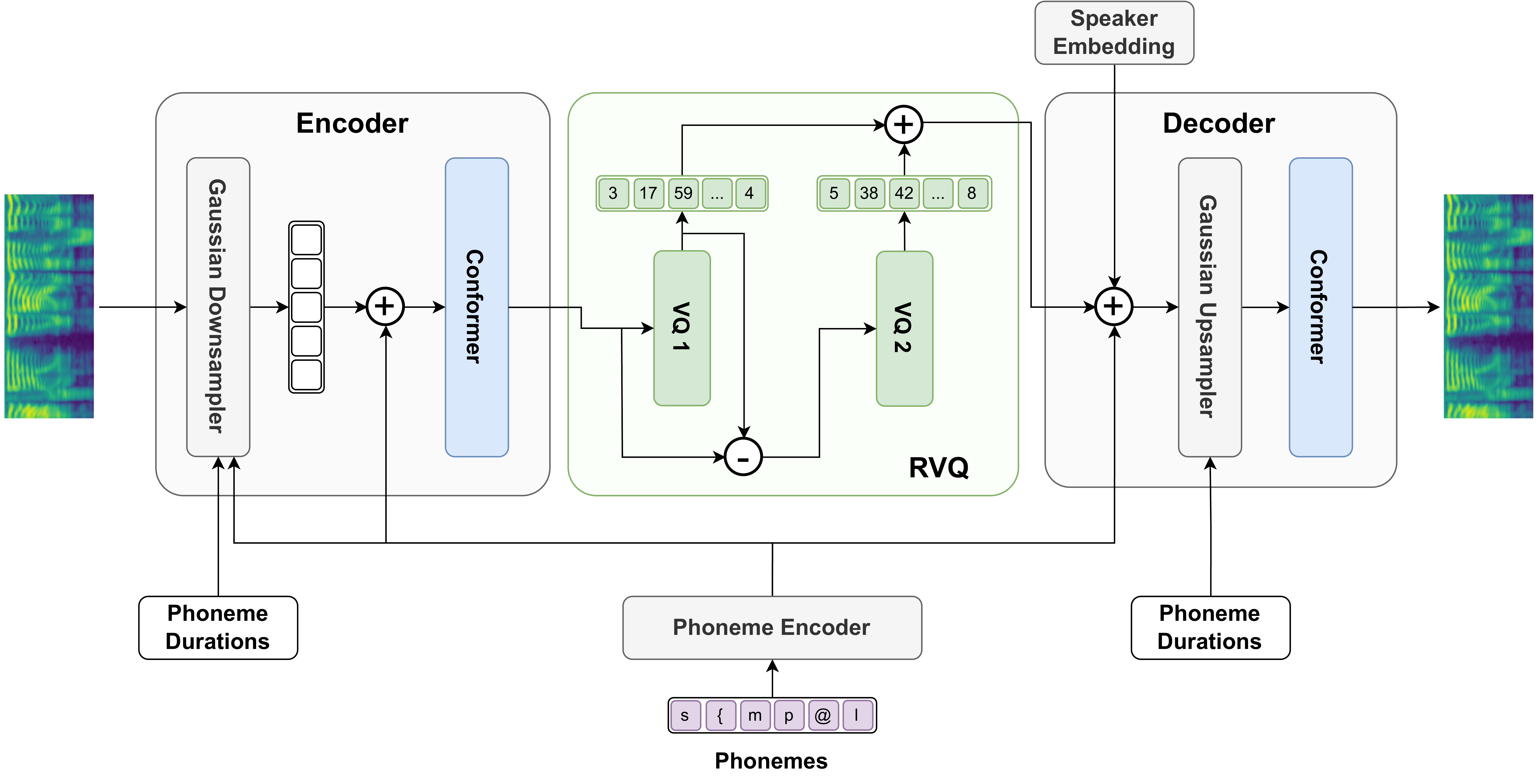}}
	\centerline{}\medskip
	\caption{The adapted architecture.}
	\label{fig:tsne}
\end{figure}

\section{Method}
\label{sec:method}

Neural codecs are appealing for TTS tasks due to their ability to tokenize a continuous modality (speech) into discrete units (codes).
After the tokenization, the speech generation task is reduced to a token generation task and the usage of powerful LLMs is enabled with state-of-the-art results.
In accordance with this, we use an adapted phoneme-level version of the SoundStream \cite{zeghidour2021soundstream} neural codec that uses multiple levels of quantization, as also presented in Mega-TTS 2 \cite{jiang2023mega}.
Simultaneously, we apply the method of conditioning the mel encoder part on linguistic representations, which is introduced in \cite{liu2022controllable} for a phoneme-level continuous VAE.
The decoder is conditioned on a learnable speaker embedding, as is common for multi-speaker models, explaining away speaker information from the previous parts of the model.

Our model attempts to reconstruct mel spectrograms of speech utterances, while also having text and speaker identity as inputs.
The input text is converted to phoneme sequences using a proprietary front-end module.
During the encoding phase, the input phonemes are preprocessed by a phoneme encoder in order to produce the linguistic features.
Also, phoneme durations are extracted by aligning phonemes to the ground-truth audio using an HMM-based forced aligner.
The input mel spectrograms are downsampled from frame-level to phoneme-level, using the extracted phoneme durations and the Gaussian Upsampler module from \cite{shen2020non}, but with a transposed attention matrix intended for downsampling instead of upsampling, as we find that this method is more effective than simple frame averaging.
The sum of phoneme-level mels and linguistic features are given as input to the conformer layers, making the encoder conditional to the linguistic content.
The continuous phoneme-level representations produced by the mel encoder are given as input to a 2-level RVQ module that produces a phoneme-level sequence for each level.

For the reconstruction phase, we again add the linguistic features and the phoneme-level RVQ sequences and upsample them using a Gaussian Upsampler module based on the ground-truth phoneme durations.
Before upsampling, a learnable speaker embedding is also broadcast-added in order to introduce speaker information.
The resulting frame-level sequence is processed by the conformer layers of the decoder in order to produce the reconstructed mel spectrogram.

We use the Conformer architecture \cite{gulati20_interspeech} for the phoneme encoder, encoder, and decoder, which was shown to be powerful in speech modeling.
The model is trained using L1 and L2 losses between the ground-truth and predicted mel spectrograms while the codebooks are updated using the Exponential Moving Average method.
Additionally a commitment loss is applied between the mel encoder outputs and the entries of the codebooks to ensure stable training of the model.

The combination of the components above forms a neural codec tailored to speech.
The applied linguistic and speaker conditioning drives the discrete latent space to disregard the related information and focus on capturing only the attributes pertaining to prosody.
This leads the model to exhibit various interesting properties which are investigated in the following section.

\section{Experiments}
\label{sec:experiments}
\subsection{Technical Details}
We train our model on a proprietary dataset comprising 4 speakers (1 male, 3 female), amounting to a total duration of approximately 100 hours of high-quality speech, balanced between the speakers.

We use 4 Conformer layers for each module, 4 attention heads, and set the model dimension to 256.
Regarding the RVQ module, we choose 2 levels of quantization using 256 codes for each layer with code dimensionality set to 3. The selections of these values was based on literature (e.g. \cite{sun2020fully}) and on empirical studies and seems to strike a good balance between performance and complexity. Specifically regarding the latent dimensionality, no improvements were observed when increasing its value during some preliminary experiments. However, a more detailed study may be needed. 

The final model has about 20 million parameters and was trained on a single Titan RTX GPU with a 16-sample batch size until convergence, which happens at around 500k iterations.

Finally, to convert the mels generated by the model into audio we use a HiFi-GAN \cite{kong2020hifi} pretrained on the same data.

The following paragraphs in this section provide extensive results in the form of objective metrics, subjective assessments from listening tests, visualizations as well as empirical observations. In addition to these, the reader is strongly encouraged to visit the webpage accompanying the paper\footnote{https://innoetics.github.io/publications/phoneme-rvq/index.html}, which offers samples generated by the model for the different experiments.

\subsection{Latent Space Analysis}

\subsubsection{Usage of Codes}
We use the trained model to extract codes from samples constituting the 10\% of the training dataset, balanced among the 4 speakers.
Initially we calculate the usage percentage of the different latent codes which turn out to be $99.6\%$ for the codes of the first level and $99.2\%$ for the second, suggesting that the bottleneck and the capacity of the encoder and decoder are appropriate.
In agreement with \cite{zeghidour2021soundstream}, we also observe that the first level carries the main burden of the modeling while the second level adds finer details. Using only the codes of the first level during resynthesis slightly degrades the average PSNR from $14.63$ to $13.97$ which, however, is not acoustically perceptible. This suggests that adding more quantization levels to this model will only have a marginal impact on its modeling performance.

We also investigate the dependency of codes between the two levels by approximating the conditional pmfs $p_{ij}$ = $P(\text{code}_2 = j| \text{code}_1 = i)$ with their corresponding histograms, where $\text{code}_n$ means code of n-th level. The average entropy of these 256 pmfs is $4.877$. Given that the entropy of the uniform distribution over 256 codes is $5.545$, the previous results suggests that there is no strong dependency between the two levels.

\subsubsection{Assessing Speaker and Linguistics Disentanglement}
To objectively assess whether any significant amount of speaker information or linguistic information are encoded in the latent representation, we examine the conditional probability mass functions of codes with regards to speaker IDs and phoneme IDs, respectively.

For speakers identity information, we calculate the conditional probability mass functions of codes given a speaker identity $P(code|speaker)$. Examining the entropy of the distributions, which are shown in Table~\ref{tab:entropies}, we observe that the pmfs are not concentrated around specific points, which would indicate that specific speakers would tend to use specific codes.

\begin{table}[th]
	\caption{Entropies of Codebooks}
	\vspace{5pt}
	\label{tab:entropies}
	\centering
	\begin{tabular}{@{\hspace*{1mm}}lcc@{\hspace*{1mm}}}
		\toprule
		\textbf{Speaker ID} & \textbf{Entropy Level 1}  & \textbf{Entropy Level 2} \\ 
		\midrule
		\textbf{0}& 5.415 & 5.454  \\
		\textbf{1}& 5.395 & 5.465 \\
		\textbf{2}& 5.382 & 5.428 \\
		\textbf{3}& 5.404 & 5.473  \\

		\bottomrule
	\end{tabular}
\end{table}

To investigate the extent to which the latent codes contain linguistic information, we calculate in a similar way the conditional pmfs of codes given the phonemes $P(code|phoneme)$ and calculate all the pairwise distances using the symmetric KL metric. We then visualize these relationships in the 2D space by employing the t-SNE with these pre-computed distances. We colorize each phoneme according to its usage. The resulting plot is shown in Fig.~\ref{fig:tsne}.

\begin{figure}[htb]
	
	\centering
	\centerline{\includegraphics[width=0.8\linewidth]{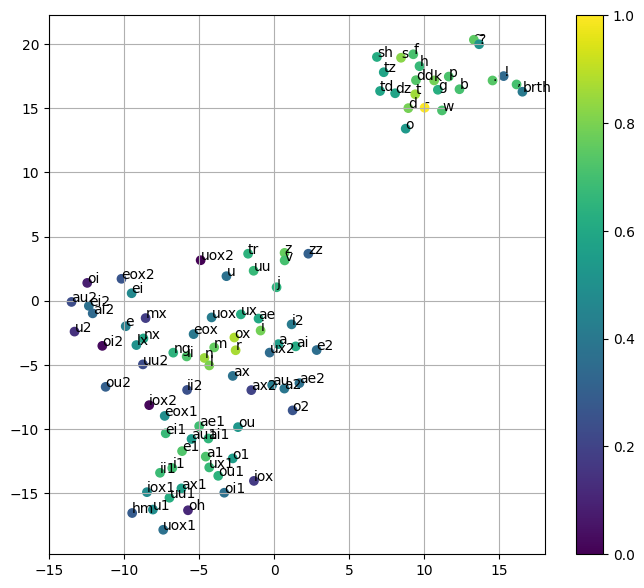}}
	\centerline{}\medskip
	\caption{Visualization of the distances between the histograms of each phoneme.}
	\label{fig:tsne}
\end{figure}

We observe that the structure of this space seems to somehow reflect the manner of articulation of the respective phonemes, with the vowels and voiced consonants forming a cluster at the bottom-left and the unvoiced consonants clustering at the top-right. Some finer structure may also be present, such as a group of nasals near the center of the voiced cluster. However, this arrangement does not seem to be too rigid, with voiced sounds appearing in the unvoiced cluster and vice versa. Moreover, phonemes that are acoustically more similar do not necessarily appear closer in this space. This may suggest that the latent representation is loosely related to the articulatory characteristics of the different phonemes but not to the phonemes' identity per se. This is also supported by the entropy values of the conditional pmfs of codes given the phonemes, as discussed above, as well as by other experiments in the following paragraphs which clearly demonstrate that latent codes may be freely combined with different phonemes without affecting their phonetic quality.

\subsubsection{Interpreting Principal Components}
\begin{figure}[htb]
	
	\centering
	\centerline{\includegraphics[width=0.8\linewidth]{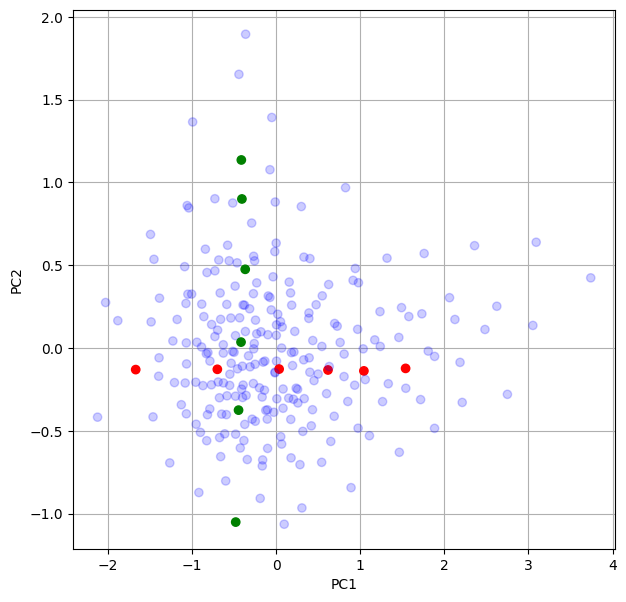}}
	\centerline{}\medskip
	\caption{Selected paths in the latent space}
	\label{fig:pca_paths}
\end{figure}

We perform Principal Components Analysis on the codes and examine the resulting components. The first two components are found to capture the vast majority of the variance of the data, namely 96\%. Fig.~\ref{fig:pca_paths} shows the codes in the 2D PCA space.

To derive a qualitative interpretation of each of those two components, we select a set of codes from that space by varying each of the two dimensions separately. To do that, we draw a path along each dimension and select codes that lie approximately on this path at different positions, as shown by the red (PCA components 1) and green (PCA components 2) dots. We then use each selected code to synthesize \textit{all} the phonemes of a reference utterance with two different speakers and analyze the generated audio for its mean pitch (F0) and mean energy (RMS). Interestingly enough, varying the first PCA dimension was highly correlated with the pitch of the generated sentence, while varying the second dimension was highly correlated with its energy, as shown in Tables~\ref{tab:pc1} and \ref{tab:pc2}.

\begin{table}[th]
	\caption{First Principal Component Interpretation}
	\vspace{5pt}
	\label{tab:pc1}
	\centering
	\begin{tabular}{@{\hspace*{1mm}}lcccc@{\hspace*{1mm}}}
		\toprule
		\textbf{Codes} & \textbf{F0} & \textbf{RMS} & \textbf{PC1} & \textbf{PC2} \\ 
		\midrule
		\textbf{106}& 90 & 0.087 & -1.669 & -0.130 \\
		\textbf{166}& 103 & 0.091 & -0.698 & -0.128 \\
		\textbf{162}& 124 & 0.095 & 0.036 & -0.127 \\
		\textbf{236}& 143 & 0.097 & 0.617 & -0.135 \\
		\textbf{12} & 157 & 0.092 & 1.043 & -0.139\\
		\textbf{107}& 172 & 0.094 & 1.539 & -0.123\\
		\bottomrule
	\end{tabular}

\end{table}
\begin{table}[th]
	\caption{Second Principal Component Interpretation}
	\vspace{5pt}
	\label{tab:pc2}
	\centering
	\begin{tabular}{@{\hspace*{1mm}}lcccc@{\hspace*{1mm}}}
		\toprule
		\textbf{Codes} & \textbf{F0} & \textbf{RMS} & \textbf{PC1} & \textbf{PC2} \\ 
		\midrule
		\textbf{15}& 111 & 0.029 & -0.413 & 1.136 \\
		\textbf{64}& 111 & 0.036 & -0.406 & 0.900 \\
		\textbf{189}& 113 & 0.052 & -0.367 & 0.475 \\
		\textbf{249}& 111 & 0.078 & -0.417 & 0.036 \\
		\textbf{86} & 110 & 0.111 & -0.446 & -0.375\\
		\textbf{184}& 109 & 0.182 & -0.480 & -0.480\\
		\bottomrule
	\end{tabular}

\end{table}

More specifically, while traversing the first path, the pitch of the generated speech was increasing when using codes with an increasing first principal component while the energy of the signal remained approximately constant. Regarding the second path, the energy of speech was increasing when using codes with a decreasing second principal component while the pitch remained approximately constant. These observations suggest that the two most significant Principal Components of the latent space model the pitch and energy of the signal. Our findings are inline with other works from the literature e.g. \cite{sun2020fully}.

We repeat the experiment with the codes from the first path for a different speaker and present the results of F0 in Table~\ref{tab:spk_f0}. As shown, inferencing with a different speaker embedding results in different but also increasing pitch values. The findings suggest that codes do not model explicitly the pitch and energy, but they rather do it in a speaker-relative way. This is a very interesting emerging property of the latent representation which makes it significantly more robust while making the latent codes highly interpretable and directly transferable among the speakers.

\begin{table}[th]
	\caption{Speaker-relative F0 results.}
	\vspace{5pt}
	\label{tab:spk_f0}
	\centering
	\begin{tabular}{@{\hspace*{1mm}}lcc@{\hspace*{1mm}}}
		\toprule
		\textbf{Codes} & \textbf{spk0} & \textbf{spk1} \\ 
		\midrule
		\textbf{106}& 90 &  160\\
		\textbf{166}& 103 &  183\\
		\textbf{162}& 124 &  211\\
		\textbf{236}& 143 &  236\\
		\textbf{12} & 157 & 261\\
		\textbf{107}& 172 & 288\\
		\bottomrule
	\end{tabular}
	
\end{table}

\subsection{Task-based Evaluation}
Following the concept of task-based evaluation employed in \cite{williams2021exploring}, we implement a number of proof-of-concept tasks to explore the properties and evaluate the performance of the proposed model by probing the model conditioning modules.
The tasks employed are meant to further assess that the speaker identity and linguistic information is explained away from the latent space, leaving the latent space to capture only prosodic attributes.

We also test the prosody transfer capabilities of the representation. We do that by using the codes derived from a reference utterance to synthesize the phonemes of a validation utterance of the same length, unseen by the model during training.

Samples from the different tasks are available in the webpage of the paper.


\subsubsection{Intelligibility - Linguistics Explained Away}
In order to assert that the linguistic content is explained away from the latent representation, we resynthesize a set of evaluation utterances using their original linguistic content as conditioning, but for each utterance we shuffle its codes before feeding them to the decoder. If any linguistic information had leaked into the representation, then this shuffling of the latent codes would tamper the identity of generated phonemes and would result in reduced intelligibility and, thus, increased ASR error rates.
As baselines for comparison we use the ground-truth samples and the plain resynthesized samples. Intelligibility is estimated by measuring the Word Error Rate (WER) and the Character Error Rate (CER) of the Whisper large Automatic Speech Recognition (ASR) pretrained model \cite{radford2023robust}.

\begin{table}[th]
	\caption{ASR-based Intelligibility evaluation}
	\vspace{5pt}
	\label{tab:intelligibility}
	\centering
	\begin{tabular}{@{\hspace*{1mm}}lcc@{\hspace*{1mm}}}
		\toprule
		\textbf{} & \textbf{WER\%$\downarrow$} & \textbf{CER\%$\downarrow$} \\
		\midrule
		Ground-truth & 1.96 & 0.55 \\
		Resynthesis & 2.31 & 0.68 \\
		Permuted codes resynthesis & 3.18 & 1.01 \\
		\bottomrule
	\end{tabular}
\end{table}

We observe that, although the WER and CER of the generated speech using shuffled codes are slightly increased, their values remain very low as compared to the ground-truth and resynthesized ones. This provides a strong indication that no significant phonetic information in present in the codes. The small deterioration of the error rates may be partly attributed to the fact that the shuffling of codes (and, consequently, prosodic information) randomly between phonemes may have some negative effect on the performance of the ASR. This is also evident by listening to the samples in webpage.

\subsubsection{Cross Resynthesis - Speaker Identity Explained Away}
In order to investigate the extent to which the speaker identity is explained away in the latent representation, we conduct a cross-resynthesis experiment where we encode audio from a source speaker and attempt to re-synthesize it using a speaker embedding of a different speaker.
Our goal here is to examine whether the resulting speech is natural, with voice timbre similar to the target speaker's, and that it retains the prosodic style of the source utterance; in other words, that there is no major degradation in the quality when using codes extracted from a specific speaker in combination with a different speaker embedding.

Naturalness evaluations were collected via a Mean Opinion Score (MOS) listening test conducted online using Amazon Mechanical Turk, in three English locales: United States (US), United Kingdom (GB) and Canada (CA).
Listeners were asked to rate the naturalness of each sample on a 5-point Likert scale, with ``1" indicating very unnatural speech and ``5" indicating completely natural speech.
Each test page contained ground-truth samples and resynthesized samples of a single speaker and cross-resynthesized samples of the same target speaker, and was rated by at least 10 unique listeners.
In total, 240 samples were evaluated, and 86 listeners participated in the listening test.

\begin{table}[th]
	\caption{Naturalness MOS evaluation}
	\vspace{5pt}
	\label{tab:naturalness}
	\centering
	\begin{tabular}{@{\hspace*{1mm}}lc@{\hspace*{1mm}}}
		\toprule
		\textbf{} & \textbf{MOS $\uparrow$} \\
		\midrule
		Ground-truth & 4.50 \scriptsize{$\pm$ 0.16} \\
		Resynthesis & 4.39 \scriptsize{$\pm$ 0.17} \\
		Cross-resynthesis & 4.28 \scriptsize{$\pm$ 0.10} \\
		\bottomrule
	\end{tabular}
\end{table}

As presented in Table~\ref{tab:naturalness}, the cross-resynthesized speech samples achieve a Mean Opinion Score on par with the resynthesized speech samples.

This clearly demonstrates two important properties of the proposed model: firstly, that the speaker characteristics are effectively absent from the latent representation and, secondly, that the prosodic information (in the form of codes) and speaker characteristics (in the form of speaker embeddings) can be freely combined to generate highly natural speech.

In addition to naturalness, we use an objective measure to quantify the similarity of the cross-resynthesized audio to the target speaker. More specifically, we calculate the similarity of the cross-resynthesized audio from various source speakers to a target speaker with the original recordings of the target speaker using d-vectors \cite{variani2014deep, wan2018generalized}, which derive a high-level representation of voice characteristics in a fixed-size vector.
For this purpose, we use Resemblyzer\footnote{https://github.com/resemble-ai/Resemblyzer}, a public pretrained d-vector encoder model.

\begin{table}[th]
	\caption{Speaker Similarity}
	\vspace{5pt}
	\label{tab:similarity}
	\centering
	\begin{tabular}{@{\hspace*{1mm}}lcc@{\hspace*{1mm}}}
		\toprule
		& \multicolumn{2}{c}{\textbf{d-vector cosine similarity $\uparrow$}} \\
		\midrule
		\textbf{} & Resynthesis & Cross-Resynthesis \\
		\midrule
		GT Spk0 & 0.986 & 0.974 \\
		GT Spk1 & 0.990 & 0.949 \\
		GT Spk2 & 0.988 & 0.977 \\
		GT Spk3 & 0.979 & 0.960 \\
		\bottomrule
	\end{tabular}
\end{table}

As shown in Table \ref{tab:similarity}, the similarity of the cross-resynthesized speech to the ground-truth samples is reduced by a negligible amount as compared to the similarity of the resynthesized samples to the ground-truth.

\begin{figure}[htb]
	\centering
	\subfloat[a][Pitch contours]{\includegraphics[width=0.8\linewidth]{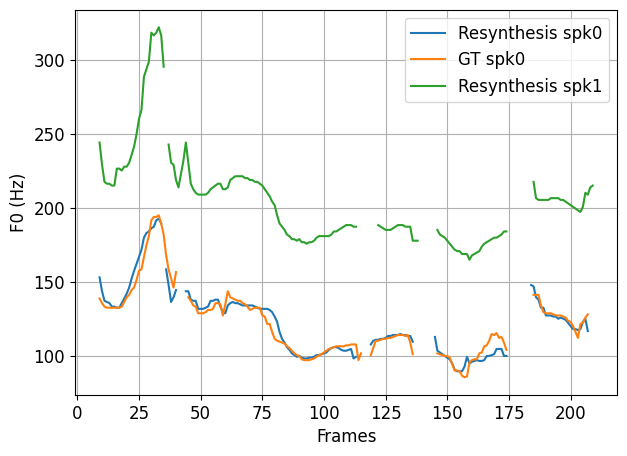} \label{subfig:contour-a}} \\
	\subfloat[b][Normalized pitch contours]{\includegraphics[width=0.8\linewidth]{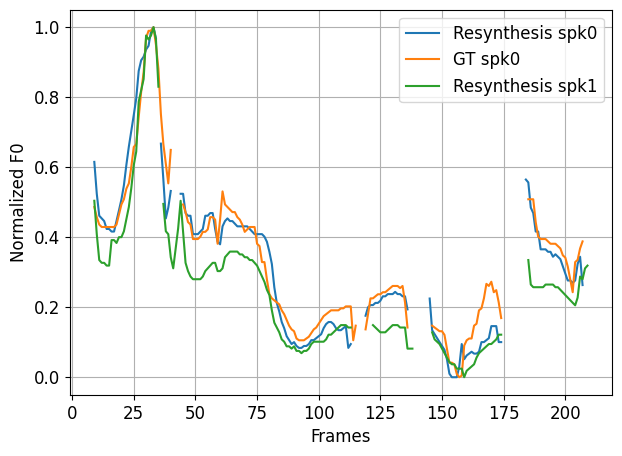} \label{subfig:contour-b}}
    \caption{Pitch contours of a ground-truth, resynthesized and cross-resynthesized speech utterance}
    \label{fig:pitch-contours}
\end{figure}

Finally, we extract and inspect the pitch contours of the ground-truth, the resynthesized and the cross-resynthesized signals (Figure \ref{subfig:contour-a}) as a way to assess the ``prosody transfer" abilities of the model, beyond speaker identity.
We also normalize the curves to remove speaker-related offsets in values (Figure \ref{subfig:contour-b}).
We observe that the model is able to encode the pitch variations of the source speech and produce speech with similar pitch variation in the target speaker's pitch range.

%
	%

%

\subsubsection{Copy Synthesis - Transferability of Latent Codes}
As a last task, we investigate the ability of the prosodic representations to be transferred between two utterances. Given that the model operates on phoneme-level features, we select a set of utterances with a specific length in phonemes. We use one of them as the target utterance while the rest of them are used as source utterances. So the goal here is to transfer the prosody from all the source utterances into the target utterance by extracting the codes from each source utterance and combining them with the phoneme embeddings of the target utterance. During the encoding phase the encoder conditions on the source text and produces the codes corresponding to the source spectrogram. During the decoding phase the decoder is conditioned on the target utterance, combines the phonemes with the previously predicted codes and estimates the mel spectrogram for the target sentence using the speaker embedding of the source speaker. In both the encoding and decoding the model uses the corresponding ground-truth durations.

We examine the results to find out if the model produces speech that corresponds to the target utterance and to what extent the prosodic attributes of the source utterance appear to the resulting speech. Using an ASR model we calculate a WER and CER of $2.05\%$ and $0.77\%$ respectively, suggesting that the synthesized speech corresponds to the target utterance. In addition, the Pearson correlation coefficient of F0 and energy between the source and target speech samples is $0.85$ and $0.45$ respectively, indicating that even with this simple experiment, prosodic features can be transferred from the source speech.

\subsection{Comparison with Continuous Model}
As described in a previous section, the codes usage percentage is in the order of $99\%$ suggesting that the discrete bottleneck forces the model to use every code available in order to model the data. In order to investigate whether the discrete bottleneck reduces the modeling capacity of the model and to what extent, we perform an ablation experiment in which we compare the aforementioned model and its \emph{continuous} latent space counterpart. More specifically, we keep the same architecture but remove the RVQ module. We evaluate the reconstruction abilities of the model using MCD, VDE, GPE, and FFE \cite{kubichek1993mel, nakatani2008method, chu2009reducing} and the results are presented in Table~\ref{tab:cont}.
We observe that the results of the continuous and discrete models are quite close indicating that, for the given task, the modeling capacity of the discrete model is fully adequate. Surprisingly, the discrete model scores slightly better, which may appear counter-intuitive since it is significantly more constrained than the continuous model. Nevertheless, it seems that this tighter bottleneck does not harm the modeling capabilities of the system in this case; in contrary, it may have a slightly beneficial role.

\begin{table}[th]
	\caption{Continuous vs Discrete Model}
	\vspace{5pt}
	\label{tab:cont}
	\centering
	\begin{tabular}{@{\hspace*{1mm}}lcccc@{\hspace*{1mm}}}
		\toprule
		\textbf{Type} & \textbf{MCD  $\downarrow$}  & \textbf{VDE $\downarrow$} & \textbf{GPE $\downarrow$} & \textbf{FFE $\downarrow$} \\ 
		\midrule
		\textbf{Discrete}& 3.49 & 11.06 & 0.76 & 11.57 \\
		\textbf{Continuous}& 3.73 & 11.40 & 1.0 & 12.04 \\

		\bottomrule
	\end{tabular}
\end{table}

\section{Conclusions \& Future Work}
\label{sec:conclusions}

We proposed a multi-speaker, phoneme-level, discrete neural codec model tailored to speech. It is specifically trained to model prosodic information in a way that is disentangled from linguistic content and from speaker characteristics, primarily capturing the prosodic aspects of the reference speech.

We performed a wide range of experiments which verified this disentanglement through objective and subjective tests.
We explored the latent representation space which proved to have some very interesting emergent properties, including the clearly interpretable semantics of its first two principal components. These turned out to correspond to pitch and energy and are encoded not as mere, absolute values but as relative quantities within the range of each speaker. These characteristics make the representation robust and controllable. Furthermore, we showed that it is at least as good as its continuous counterpart, based on several objective metrics.

The performance of the model was assessed on various tasks, including cross-synthesis and prosody transfer, with the results further supporting the robustness of the representation. 
Despite its very modest size and complexity the model was able to generate speech of very high naturalness when varying linguistic input, prosodic codes, and/or speaker information separately. This remained true when the latent codes where changed arbitrarily and even when they were shuffled randomly.

As part of future research, it would be quite interesting to examine the quality of the representation and the overall performance of the system when trained on much larger datasets comprising hundreds or thousands of speakers. Integrating a duration predictor into the model will also offer some benefits, as it will alleviate from the need to rely on fixed, pre-calculated reference durations.
But a really interesting hypothesis to test in the future would be whether the robust latent representation obtained this way is indeed more amenable to prediction and control; i.e. whether such representation makes it more efficient to train effective prior models to predict latent codes directly from phonemes and/or other control signals, as part of a TTS or a voice conversion system.

\bibliographystyle{IEEEbib}
\bibliography{refs}

\end{document}